# Internet of Things (IoT): A Vision, Architectural Elements, and Future Directions

Jayavardhana Gubbi,[a] Rajkumar Buyya,[b*] Slaven Marusic,[a] Marimuthu Palaniswami[a]

[a]*Department of Electrical and Electronic Engineering, The University of Melbourne, Vic - 3010, Australia*

[b]*Department of Computing and Information Systems, The University of Melbourne, Vic - 3010, Australia*

**Abstract**

Ubiquitous sensing enabled by Wireless Sensor Network (WSN) technologies cuts across many areas of modern day living. This offers the ability to measure, infer and understand environmental indicators, from delicate ecologies and natural resources to urban environments. The proliferation of these devices in a communicating-actuating network creates the Internet of Things (IoT), wherein, sensors and actuators blend seamlessly with the environment around us, and the information is shared across platforms in order to develop a common operating picture (COP). Fuelled by the recent adaptation of a variety of enabling device technologies such as RFID tags and readers, near field communication (NFC) devices and embedded sensor and actuator nodes, the IoT has stepped out of its infancy and is the the next revolutionary technology in transforming the Internet into a fully integrated Future Internet. As we move from www (static pages web) to web2 (social networking web) to web3 (ubiquitous computing web), the need for data-on-demand using sophisticated intuitive queries increases significantly. This paper presents a cloud centric vision for worldwide implementation of Internet of Things. The key enabling technologies and application domains that are likely to drive IoT research in the near future are discussed. A cloud implementation using *Aneka,* which is based on interaction of private and public clouds is presented. We conclude our IoT vision by expanding on the need for convergence of WSN, the Internet and distributed computing directed at technological research community.

## 1. Introduction

The next wave in the era of computing will be outside the realm of the traditional desktop. In the Internet of Things (IoT) paradigm, many of the objects that surround us will be on the network in one form or another. Radio Frequency IDentification (RFID) and sensor network technologies will rise to meet this new challenge, in which information and communication systems are invisibly embedded in the environment around us. This results in the generation of enormous amounts of data which have to be stored, processed and presented in a seamless, efficient and easily interpretable form. This model will consist of services that are commodities and delivered in a manner similar to traditional commodities. Cloud computing can provide the virtual infrastructure for such utility computing which integrates monitoring devices, storage devices, analytics tools, visualization platforms and client delivery. The cost based model that Cloud computing offers will enable end-to-end service provisioning for businesses and users to access applications on demand from anywhere.

Smart connectivity with existing networks and context-aware computation using network resources is an indispensable part of IoT. With the growing presence of WiFi and 4G-LTE wireless Internet access, the evolution toward ubiquitous information and communication networks is already evident. However, for the Internet of Things vision to successfully emerge, the computing criterion will need to go beyond traditional mobile computing scenarios that use smart phones and portables, and evolve into connecting everyday existing objects and embedding intelligence into our environment. For technology to *disappear* from the consciousness of the user, the Internet of Things demands: (1) a shared understanding of the situation of its users and their appliances, (2) software architectures and pervasive communication networks to process and convey the contextual information to where it is relevant, and (3) the analytics tools in the Internet of Things that aim for autonomous and smart behavior. With these three fundamental grounds in place, smart connectivity and context-aware computation can be accomplished.

[*] Corresponding author. Tel.: +61 3 83441344; fax: +61 3 93481184; e-mail:rbuyya@unimelb.edu.au; url:www.buyya.com.

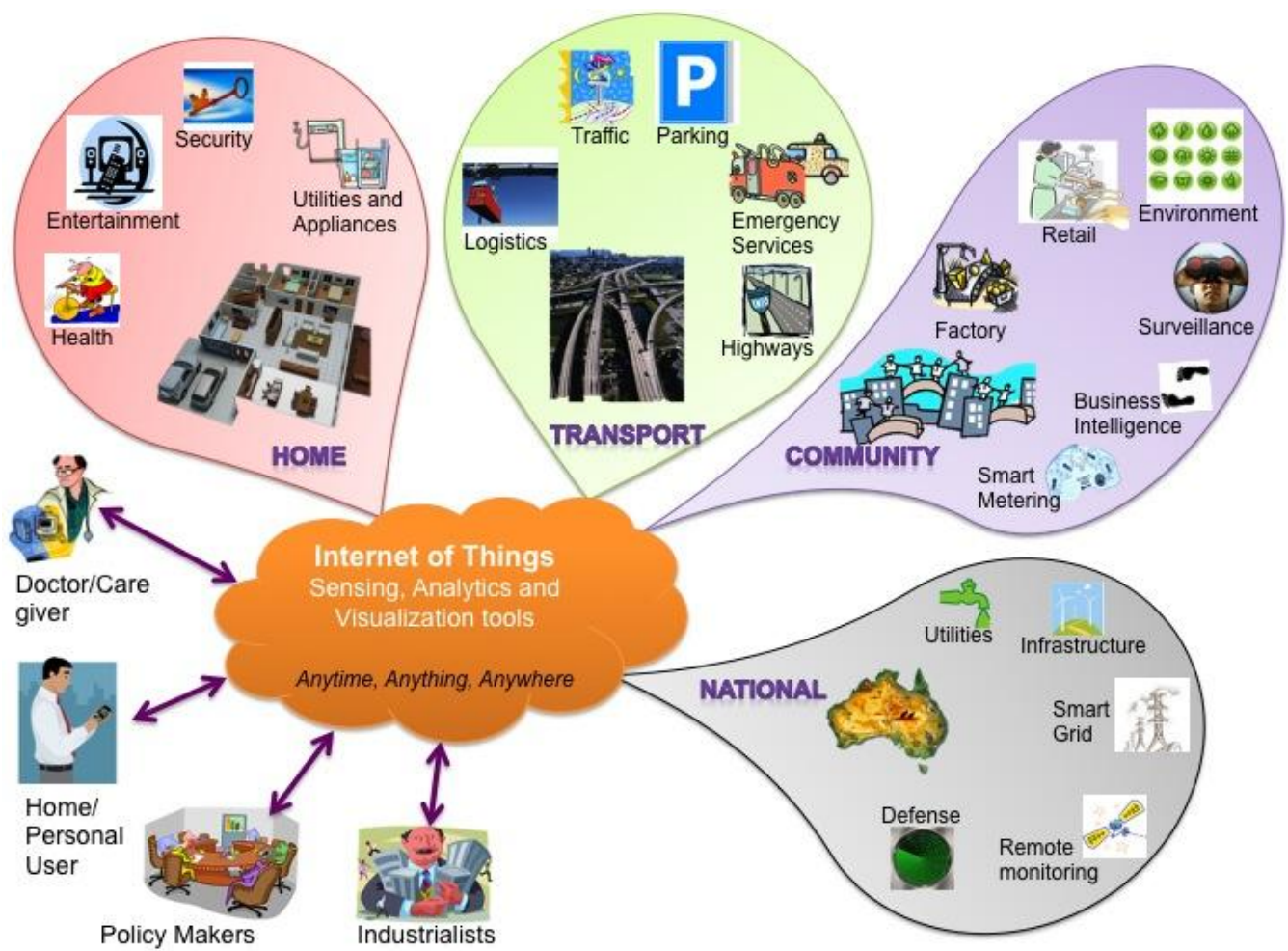

Figure 1: Internet of Things Schematic showing the end users and application areas based on data

A radical evolution of the current Internet into a Network of interconnected *objects* that not only harvests information from the environment (sensing) and interacts with the physical world (actuation/command/control), but also uses existing Internet standards to provide services for information transfer, analytics, applications and communications. Fuelled by the prevalence of devices enabled by open wireless technology such as Bluetooth, radio frequency identification (RFID), Wi-Fi and telephonic data services as well as embedded sensor and actuator nodes, IoT has stepped out of its infancy and is on the verge of transforming the current static Internet into a fully integrated Future Internet [1]. Internet revolution led to the interconnection between people at an unprecedented scale and pace. The next revolution will be the interconnection between objects to create a smart environment. Only in 2011, the number of interconnected devices on the planet overtook the actual number of people. Currently there are 9 billion interconnected devices and it is expected to reach 24 billion devices by 2020. According to the GSMA, this amounts to $1.3 trillion revenue opportunities for mobile network operators alone spanning vertical segments such as health, automotive, utilities and consumer electronics. A schematic of the interconnection of objects is depicted in Figure 1 where the application domains are chosen based on the scale of the impact of the data generated. The users span from an individual to national level organizations addressing wide ranging issues.

This paper presents the current trends in IoT research propelled by applications and the need for convergence in several interdisciplinary technologies. Specifically, we present:

- Overall IoT vision and the technologies that will achieve the it (Section 2)
- Some common definitions in the area along with some trends and taxonomy of IoT (Section 3)
- Application domains in IoT with a new approach in defining them (Section 4)
- Cloud centric IoT realization and challenges (Section 5)

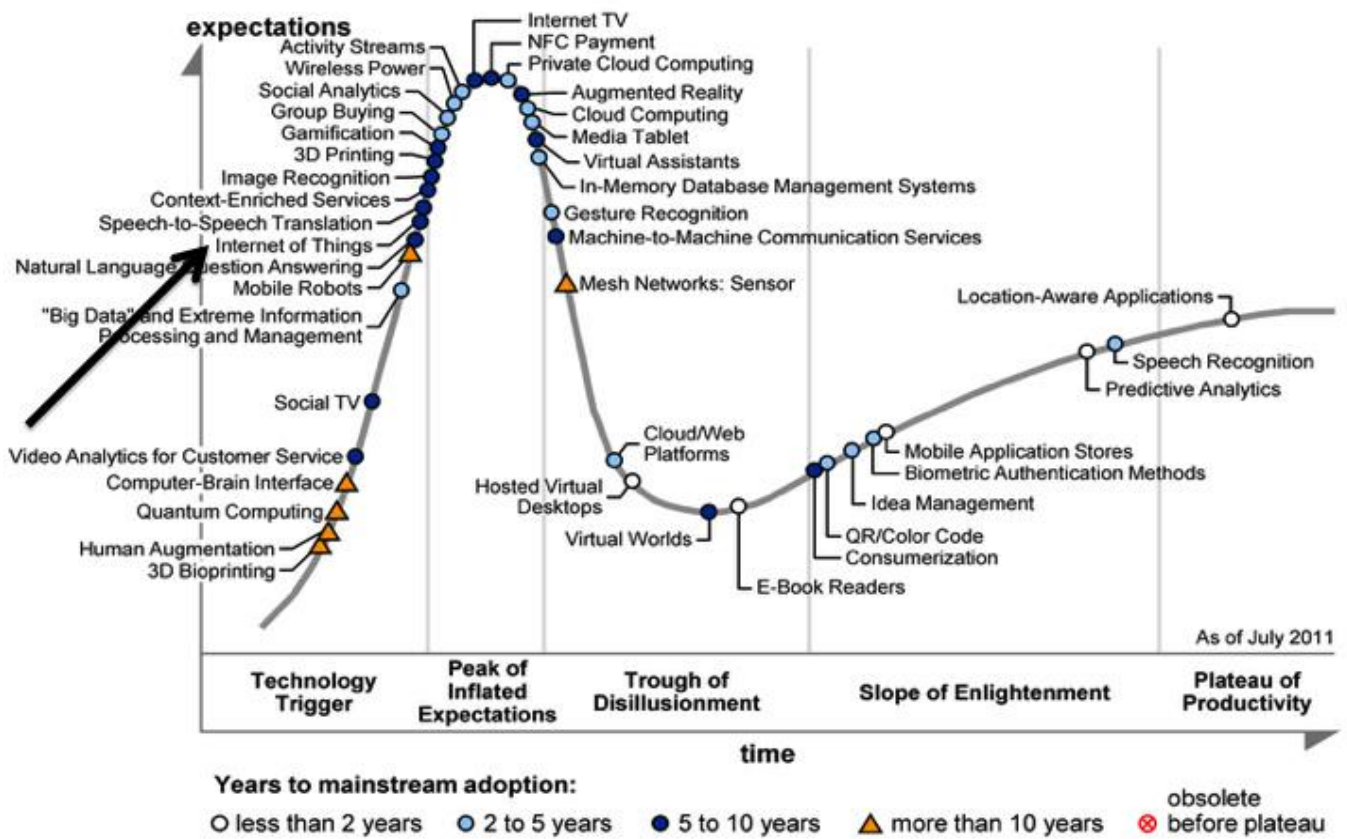

Figure 2: Gartner 2011 Hype Cycle of Emerging Technologies (Source: Gartner Inc. [10])

- Case study of data analytics on the Aneka/Azure cloud platform (Section 6)
- Open challenges and future trends (Section 7)

## 2. Ubiquitous computing in the next decade

The effort by researchers to create human-to-human interface through technology in the late 1980s resulted in the creation of the ubiquitous computing discipline, whose objective is to embed technology into the background of everyday life. Currently, we are in the post-PC era where smart phones and other handheld devices are changing our environment by making it more interactive as well as informative. Mark Weiser, the forefather of Ubiquitous Computing (ubicomp), defined a smart environment [2] as "the physical world that is richly and invisibly interwoven with sensors, actuators, displays, and computational elements, embedded seamlessly in the everyday objects of our lives, and connected through a continuous network."

The creation of the Internet has marked a foremost milestone towards achieving ubicomp's vision which enables individual devices to communicate with any other device in the world. The inter-networking reveals the potential of a seemingly endless amount of distributed computing resources and storage owned by various owners.

In contrast to Weiser's calm computing approach, Rogers proposes a human centric ubicomp which makes use of human creativity in exploiting the environment and extending their capabilities [3]. He proposes a domain specific ubicomp solution when he says –"In terms of who should benefit, it is useful to think of how ubicomp technologies can be developed not for the Sal's of the world, but for particular domains that can be set up and customized by an individual firm or organization, such as for agriculture production, environmental restoration or retailing."

Caceres and Friday [4] discuss the progress, opportunities and challenges during the 20 year anniversary of ubicomp. They discuss the building blocks of ubicomp and the characteristics of the system to adapt to the changing world. More importantly, they identify two critical technologies for growing the ubicomp infrastructure - *Cloud Computing* and the *Internet of Things*.

The advancements and convergence of micro-electro-mechanical systems (MEMS) technology, wireless communications and digital electronics has resulted in the development of miniature devices having the ability to sense, compute and communicate wirelessly in short distances. These miniature devices called nodes interconnect to form a wireless sensor networks (WSN) and find wide application in environmental monitoring, infrastructure monitoring, traffic monitoring, retail, etc. [5].

This has the ability to provide ubiquitous sensing capability which is critical in realizing the overall vision of ubicomp as outlined by Weiser [2].

For the realization of a complete IoT vision, an efficient, secure, scalable and market oriented computing and storage resourcing is essential. Cloud computing [6] is the most recent paradigm to emerge which promises reliable services delivered through next generation data centres that are based on virtualised storage technologies. This platform acts as a receiver of data from the ubiquitous sensors; as a computer to analyze and interpret the data; as well as providing the user with easy to understand web based visualization. The ubiquitous sensing and processing works in the background, *hidden* from the user.

This novel integrated Sensor-Actuator-Internet framework shall form the core technology around which a smart environment will be shaped: information generated will be shared across diverse platforms and applications, to develop a common operating picture (COP) of an environment, where control of certain unrestricted 'Things' is made possible. As we move from www (static pages web) to web2 (social networking web) to web3 (ubiquitous computing web), the need for data-on-demand using sophisticated intuitive queries increases. To take full advantage of the available Internet technology, there is a need to deploy large-scale, platform-independent, wireless sensor network infrastructure that includes data management and processing, actuation and analytics. Cloud computing promises high reliability, scalability and autonomy to provide ubiquitous access, dynamic resource discovery and composability required for the next generation Internet of Things applications. Consumers will be able to choose the service level by changing the Quality of Service parameters.

**3. Definitions, Trends and Elements**

*3.1. Definitions*

As identified by Atzori *et. al.* [7], Internet of Things can be realized in three paradigms – internet-oriented (middleware), things oriented (sensors) and semantic-oriented (knowledge). Although this type of delineation is required due to the interdisciplinary nature of the subject, the usefulness of IoT can be unleashed only in an application domain where the three paradigms intersect.

The RFID group defines Internet of Things as –
- The worldwide network of interconnected objects uniquely addressable based on standard communication protocols.

According to Cluster of European research projects on the Internet of Things [8] –

- 'Things' are active participants in business, information and social processes where they are enabled to interact and communicate among themselves and with the environment by exchanging data and information sensed about the environment, while reacting autonomously to the real/physical world events and influencing it by running processes that trigger actions and create services with or without direct human intervention.

According to Forrester [9], a smart environment –
- Uses information and communications technologies to make the critical infrastructure components and services of a city administration, education, healthcare, public safety, real estate, transportation and utilities more aware, interactive and efficient.

In our definition, we make the definition more user centric and do not restrict it to any standard communication protocol. This will allow long-lasting applications to be developed and deployed using the available state-of-the-art protocols at any given point in time. Our definition of Internet of Things for smart environments is –
- Interconnection of sensing and actuating devices providing the ability to share information across platforms through a unified framework, developing a common operating picture for enabling innovative applications. This is achieved by seamless large scale sensing, data analytics and information representation using cutting edge ubiquitous sensing and cloud computing.

*3.2. Trends*

Internet of Things has been identified as one of the emerging technologies in IT as noted in Gartner's IT Hype Cycle (see Figure 2). A Hype Cycle [10] is a way to represent the emergence, adoption, maturity and impact on applications of specific technologies. It has been forecasted that IoT will take 5-10 years for market adoption.

The popularity of different paradigms varies with time. The web search popularity, as measured by the Google search trends during the last 10 years for the terms Internet of Things, Wireless Sensor Networks and Ubiquitous Computing are shown in Figure 3 [11]. As it can be seen, since IoT has come into existence, search volume is consistently increasing with the falling trend for Wireless Sensor Networks. This trend is likely to continue for the next decade as other enabling technologies converge to form a genuine Internet of Things. From the News Reference Volume data (see Figure 3 bottom) the Internet be observed that Internet of Things has started gaining popularity. In fact, this reflects the social acceptability of the technology as consumers look for more data about various topics of interest. Spot points in Figure 3 indicate

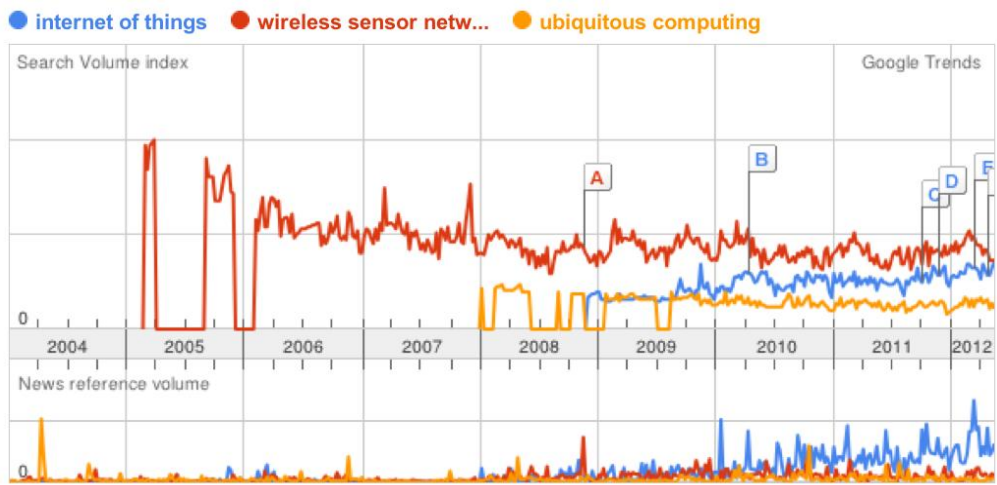

Figure 3: Google search trends since 2004: Internet of Things, Wireless Sensor Networks, Ubiquitous Computing. SPOT points are listed:

A. Algorithms and Protocols for Wireless Sensor Networks Provides You with A Comprehensive Resource MarketWatch - Nov 18 2008
B. Internet of Things -- From Vision to Reality MarketWatch - Apr 14 2010
C. CCID Consulting: China's Internet-of-Things Industry Sees a Landscape Characterized by Clustering in Four Regions MarketWatch - Oct 4 2011
D. China Hi-Tech Fair Highlights Internet of Things MarketWatch - Nov 21 2011
E. ARM unveils low-power chip for the 'internet of things' Reuters UK - Mar 13 2012
F. Web connected objects get a 'voice' on the 'Internet of Things' Winnipeg Free Press - Apr 25 2012

the news related to IoT and the highest hits are listed in Figure 3 caption.

*3.3. IoT Elements*

We present a taxonomy that will aid in defining the components required for Internet of Things from a high level perspective. Specific taxonomies of each component can be found elsewhere [6,12,13]. There are three IoT components which enables seamless ubicomp: a) Hardware - made up of sensors, actuators and embedded communication hardware b) Middleware - on demand storage and computing tools for data analytics and c) Presentation - novel easy to understand visualization and interpretation tools which can be widely accessed on different platforms and which can be designed for different applications. In this section, we discuss a few enabling technologies in these categories which will make up the three components stated above.

*3.3.1. Radio Frequency Identification (RFID)*
RFID technology is a major breakthrough in the embedded communication paradigm which enables design of microchips for wireless data communication. They help in automatic identification of anything they are attached to acting as an electronic barcode [14,15]. The passive RFID tags are not battery powered and they use the power of the reader's interrogation signal to communicate the ID to the RFID reader. This has resulted in many applications particularly in retail and supply chain management. The applications can be found in transportation (replacement of tickets, registration stickers) and access control applications as well. The passive tags are currently being used in many bank cards and road toll tags which is among the first global deployments. Active RFID readers have their own battery supply and can instantiate the communication. Of the several applications, the main application of active RFID tags is in port containers [15] for monitoring cargo.

*3.3.2. Wireless Sensor Networks (WSN)*
Recent technological advances in low power integrated circuits and wireless communications have made available efficient, low cost, low power miniature devices for use in remote sensing applications. The combination of these factors has improved the viability of utilizing a sensor network consisting of a large number of intelligent sensors, enabling the collection, processing, analysis and dissemination of valuable information, gathered in a variety of environments [5]. Active RFID is nearly the same as the lower end WSN nodes with limited processing capability and storage. The scientific challenges that must be overcome in order to realize the enormous potential of WSNs are substantial and multidisciplinary in nature [5]. Sensor data are shared among sensor nodes and sent to a distributed or centralized system for analytics. The components that make up the WSN monitoring network include:

a) WSN hardware - Typically a node (WSN core hardware) contains sensor interfaces, processing units, transceiver units and power supply. Almost always, they comprise of multiple A/D converters for sensor interfacing and more modern sensor nodes have the ability to communicate using one frequency band making them more versatile [5].
b) WSN communication stack - The nodes are expected to be deployed in an adhoc manner for most applications. Designing an appropriate topology, routing and MAC layer is critical for scalability and longevity of the deployed network. Nodes in a WSN need to communicate among themselves to transmit data in single or multi-hop to a base station. Node drop outs, and consequent degraded network lifetimes, are frequent. The communication stack at the sink node should be able to interact with the outside world through the Internet to act as a gateway to the WSN subnet and the Internet [16].
c) Middleware - A mechanism to combine cyber infrastructure with a Service Oriented Architecture (SOA) and sensor networks to provide access to heterogeneous sensor resources in a deployment independent manner [17]. This is based on the idea to isolate resources that can be used by several applications. A platform independent middleware for developing sensor applications is required, such as an Open Sensor Web Architecture (OSWA) [18]. OSWA is built upon a uniform set of operations and standard data representations as defined in the Sensor Web Enablement Method (SWE) by the Open Geospatial Consortium (OGC).
d) Secure Data aggregation - An efficient and secure data aggregation method is required for extending the lifetime of the network as well as ensuring reliable data collected from sensors [19]. Node failure being a common characteristic of WSNs, the network topology should have the capability to heal itself. Ensuring security is critical as the system is automatically linked to actuators and protecting the systems from intruders becomes very important.

*3.3.3. Addressing schemes*

The ability to uniquely identify 'Things' is critical for the success of IoT. This will not only allow us to uniquely identify billions of devices but also to control remote devices through the Internet. The few most critical features of creating a unique address are: uniqueness, reliability, persistence and scalability.

Every element that is already connected and those that are going to be connected, must be identified by their unique identification, location and functionalities. The current IPv4 may support to an extent where a group of cohabiting sensor devices can be identified geographically, but not individually. The Internet Mobility attributes in the IPV6 may alleviate some of the device identification problems; however, the heterogeneous nature of wireless nodes, variable data types, concurrent operations and confluence of data from devices exacerbates the problem further [20].

Persistent network functioning to channel the data traffic ubiquitously and relentlessly is another aspect of IoT. Although, the TCP/IP takes care of this mechanism by routing in a more reliable and efficient way, from source to destination, the IoT faces a bottleneck at the interface between the gateway and wireless sensor devices. Furthermore, the scalability of the device address of the existing network must be sustainable. The addition of networks and devices must not hamper the performance of the network, the functioning of the devices, the reliability of the data over the network or the effective use of the devices from the user interface.

To address these issues, the Uniform Resource Name (URN) system is considered fundamental for the development of IoT. URN creates replicas of the resources that can be accessed through the URL. With large amounts of spatial data being gathered, it is often quite important to take advantage of the benefits of metadata for transferring the information from a database to the user via the Internet [21]. IPv6 also gives a very good option to access the resources uniquely and remotely. Another critical development in addressing is the development of a light weight IPv6 that will enable addressing home appliances uniquely.

Wireless sensor networks (considering them as building blocks of IoT), which run on a different stack compared to the Internet, cannot possess IPv6 stack to address individually and hence a subnet with a gateway having a URN will be required. With this in mind, we then need a layer for addressing sensor devices by the relevant gateway. At the subnet level, the URN for the sensor devices could be the unique IDs rather than human-friendly names as in the www, and a lookup table at the gateway to address this device. Further, at the node level each sensor will have a URN (as numbers) for sensors to be addressed by the gateway. The entire network now forms a web of connectivity from users (high-level) to sensors (low-level) that is addressable (through URN), accessible (through URL) and controllable (through URC).

*3.3.4. Data storage and analytics*

One of the most important outcomes of this emerging field is the creation of an unprecedented amount of data. Storage, ownership and expiry of the data become critical issues. The internet consumes up to 5% of the total energy generated today and with these types of demands, it is sure to go up even further. Hence data centers which run on harvested energy and which are centralized will ensure energy efficiency as well as reliability. The data have to be stored and used intelligently for smart monitoring and actuation. It is important to develop artificial intelligence algorithms which could be centralized or distributed based on the need. Novel fusion algorithms need to be developed to make sense of the data collected. State-of-the-art non-linear, temporal machine learning methods based on evolutionary algorithms, genetic algorithms, neural networks, and other artificial intelligence techniques are necessary to achieve automated decision making. These systems show characteristics such as interoperability, integration and adaptive communications. They also have a modular architecture both in terms of hardware system design as well as software development and are usually very well-suited for IoT applications.

*3.3.5. Visualization*

Visualization is critical for an IoT application as this allows interaction of the user with the environment. With recent advances in touch screen technologies, use of smart tablets and phones has become very intuitive. For a lay person to fully benefit from the IoT revolution, attractive and easy to understand visualization have to be created. As we move from 2D to 3D screens, more information can be provided to the user in meaningful ways for consumers. This will also enable policy makers to convert data into knowledge which is critical in fast decision making. Extraction of meaningful information from raw data is non-trivial. This encompasses both event detection and visualization of the associated raw and modelled data, with information represented according to the needs of the end-user.

**4. Applications**

There are several application domains which will be impacted by the emerging Internet of Things. The applications can be classified based on the type of network availability, coverage, scale, heterogeneity, repeatability, user involvement and impact [22]. We categorize the applications into four application domains: (1) Personal and Home; (2) Enterprise; (3) Utilities; and (4) Mobile. This is depicted in Figure 1 which represents Personal and Home IoT at the scale of an individual or home, Enterprise IoT at the scale of a community, Utility IoT at a national or regional scale and Mobile IoT which is usually spread across other domains mainly due to the nature of connectivity and scale. There is a huge crossover in applications and the use of data between the domains. For instance, the Personal and Home IoT produces electricity usage data in the house and makes it available to the electricity (utility) company which can in turn optimizes the supply and demand in the Utility IoT. Internet enables sharing of data between different service providers in a seamless manner creating multiple business opportunities. A few typical applications in each domain are given.

*4.1. Personal and Home*

The sensor information collected is used only by the individuals who directly own the network. Usually WiFi is used as the backbone enabling higher bandwidth data (video) transfer as well as higher sampling rates (Sound).

Ubiquitous healthcare [7] has been envisioned for the past two decades. IoT gives a perfect platform to realize this vision using body area sensors and IoT backend to upload the data to servers. For instance, a Smartphone can be used for communication along with several interfaces like Bluetooth for interfacing sensors measuring physiological parameters. So far, there are several applications available for Apple iOS, Google Android and Windows Phone operating system that measure various parameters. However, it is yet to be centralized in the cloud for general physicians to access the same.

An extension of the personal body area network is creating a home monitoring system for aged-care, which allows the doctor to monitor patients and elderly in their homes thereby reducing hospitalization costs through early intervention and treatment [23,24].

Control of home equipment such as air conditioners, refrigerators, washing machines etc., will allow better home and energy management. This will see consumers become involved in IoT revolution in the same manner as the Internet revolution itself [25,26]. Social networking is set to undergo another transformation with billions of interconnected objects [27,28]. An interesting development will be using a Twitter like concept where individual 'Things' in the house can periodically tweet the readings which can be easily followed from anywhere creating a *TweetOT*. Although this provides a common framework using cloud for information access, a new security paradigm will be required for this to be fully realized [29].

*4.2. Enterprise*

We refer to the 'Network of Things' within a work environment as an enterprise based application. Information collected from such networks are used only by the owners and the data may be released selectively. Environmental monitoring is the first common application which is implemented to keep a track of the number of occupants and manage the utilities within the building (e.g., HVAC, lighting).

Table 1: Smart environment application domains

|  | Smart Home/Office | Smart Retail | Smart City | Smart Agriculture/Forest | Smart Water | Smart transportation |
|---|---|---|---|---|---|---|
| Network Size | Small | Small | Medium | Medium/Large | Large | Large |
| Users | Very few, family members | Few, community level | Many, policy makers, general public | Few, landowners, policy makers | Few, government | Large, general public |
| Energy | Rechargeable battery | Rechargeable battery | Rechargeable battery, Energy harvesting | Energy harvesting | Energy harvesting | Rechargeable battery, Energy harvesting |
| Internet connectivity | Wifi, 3G, 4G LTE backbone | Wifi, 3G, 4G LTE backbone | Wifi, 3G, 4G LTE backbone | Wifi, Satellite communication | Satellite Communication, Microwave links | Wifi, Satellite Communication |
| Data management | Local server | Local server | Shared server | Local server, Shared server | Shared server | Shared server |
| IoT Devices | RFID, WSN | Smart Retail | RFID, WSN | WSN | Single sensors | RFID, WSN, Single sensors |
| Bandwidth requirement | Small | Small | Large | Medium | Medium | Medium/Large |
| Example testbeds | Aware Home [31] | SAP Future retail center [32] | Smart Santander[33], CitySense [34] | SiSViA [35] | GBROOS [36], SEMAT [37] | A few trial implementations [38,39] |

Sensors have always been an integral part of factory setup for security, automation, climate control, etc. This will eventually be replaced by wireless system giving the flexibility to make changes to the setup whenever required. This is nothing but an IoT subnet dedicated to factory maintenance.

Table 2: Potential IoT applications identified by different focus groups of City of Melbourne

| **Citizens** | |
|---|---|
| Healthcare | triage, patient monitoring, personnel monitoring, disease spread modelling and containment - real-time health status and predictive information to assist practitioners in the field, or policy decisions in pandemic scenarios |
| Emergency services, defence | remote personnel monitoring (health, location); resource management and distribution, response planning; sensors built into building infrastructure to guide first responders in emergencies or disaster scenarios |
| Crowd monitoring | crowd flow monitoring for emergency management; efficient use of public and retail spaces; workflow in commercial environments |
| **Transport** | |
| Traffic management | Intelligent transportation through real-time traffic information and path optimisation |
| Infrastructure monitoring | sensors built into infrastructure to monitor structural fatigue and other maintenance; accident monitoring for incident management and emergency response coordination |
| **Services** | |
| Water | water quality, leakage, usage, distribution, waste management |
| Building management | temperature, humidity control, activity monitoring for energy usage management Ð Heating, Ventilation and Air Conditioning (HVAC) |
| Environment | Air pollution, noise monitoring, waterways, industry monitoring |

One of the major IoT application areas which is already drawing attention is Smart Environment IoT [22,29]. There are several testbeds being implemented and many more planned in the coming years. Smart environment includes subsystems as shown in Table 1 and the characteristics from a technological perspective are listed briefly. It should be noted that each of the sub domains cover many focus groups and the data will be shared. The applications or use-cases within the urban environment that can benefit from the realisation of a smart city WSN capability are shown in Table 2. These applications are grouped according to their impact areas. This includes the effect on citizens considering health and well being issues; transport in light of its impact on mobility, productivity, pollution; and services in terms of critical community services managed and provided by local government to city inhabitants.

*4.3. Utilities*

The information from the networks in this application domain are usually for service optimisation rather than consumer consumption. It is already being used by utility companies (smart meter by electricity supply companies) for resource management in order to optimise cost *vs.* profit. These are made up of very extensive networks (usually laid out by large organisation on regional and national scale) for monitoring critical utilities and efficient resource management. The backbone network used can vary between cellular, WiFi and satellite communication.

Smart grid and smart metering is another potential IoT application which is being implemented around the world [40]. Efficient energy consumption can be achieved by continuously monitoring every electricity point within a house and using this information to modify the way electricity is consumed. This information at the city scale is used for maintaining the load balance within the grid ensuring high quality of service.

Video based IoT [41] which integrates image processing, computer vision and networking frameworks will help develop a new challenging scientific research area at the intersection of video, infrared, microphone and network technologies. Surveillance, the most widely used camera network applications, helps track targets, identify suspicious activities, detect left luggage and monitor unauthorized access. Automatic behavior analysis and event detection (as part of sophisticated video analytics) is in its infancy and breakthroughs are expected in the next decade as pointed out in the 2011 Gartner Chart (refer Figure 2) congestion causing freight delays and delivery schedule failures. Dynamic traffic information will affect freight movement, allow better planning and improved scheduling. The transport IoT will enable the use of large scale WSNs

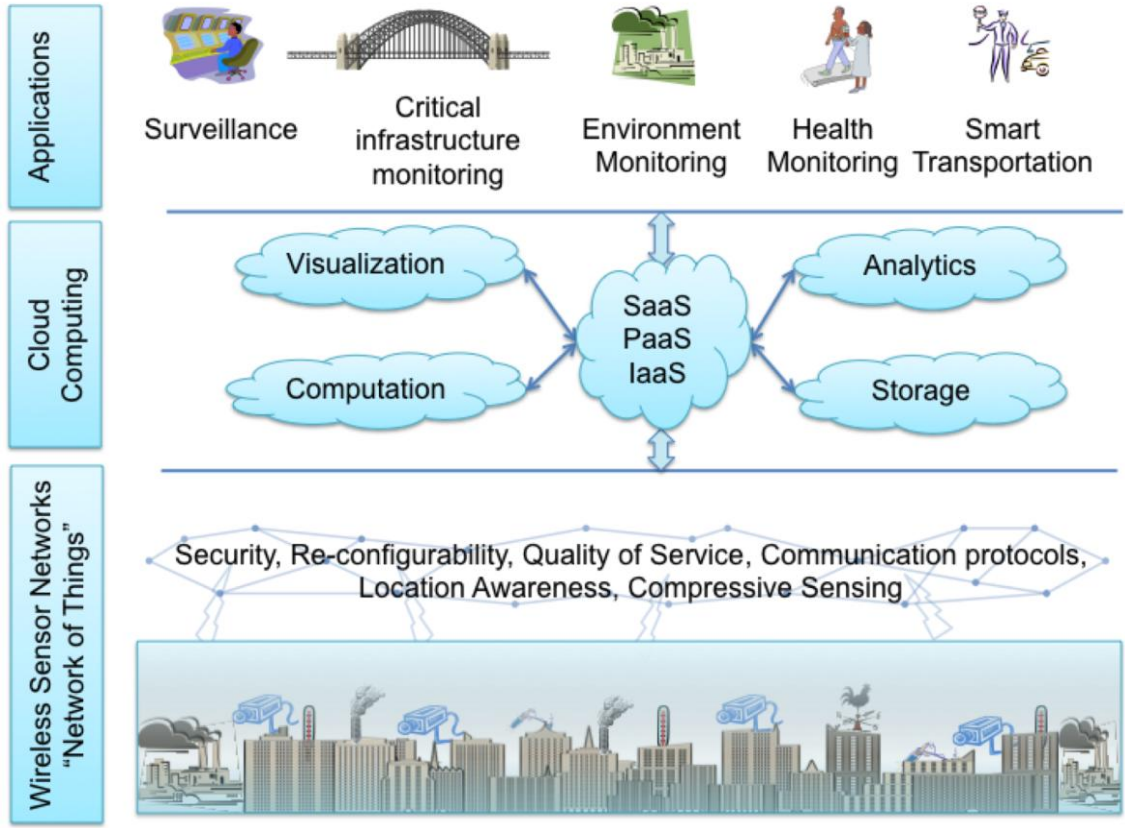

Figure 4: Conceptual IoT framework with Cloud Computing at the centre

Water network monitoring and quality assurance of drinking water is another critical application that is being addressed using IoT. Sensors measuring critical water parameters are installed at important locations in order to ensure high supply quality. This avoids accidental contamination among storm water drains, drinking water and sewage disposal. The same network can be extended to monitor irrigation in agricultural land. The network is also extended for monitoring soil parameters which allows informed decision making about agriculture [42].

*4.4. Mobile*

Smart transportation and smart logistics are placed in a separate domain due to the nature of data sharing and backbone implementation required. Urban traffic is the main contributor to traffic noise pollution and a major contributor to urban air quality degradation and greenhouse gas emissions. Traffic congestion directly imposes significant costs on economic and social activities in most cities. Supply chain efficiencies and productivity, including just-in-time operations, are severely impacted by this for online monitoring of travel times, origin-destination (O-D) route choice behavior, queue lengths and air pollutant and noise emissions. The IoT is likely to replace the traffic information provided by the existing sensor networks of inductive loop vehicle detectors employed at the intersections of existing traffic control systems. They will also underpin the development of scenario-based models for planning and design of mitigation and alleviation plans, as well as improved algorithms for urban traffic control, including multi-objective control systems. Combined with information gathered from the urban traffic control system, valid and relevant information on traffic conditions can be presented to travelers [43].

The prevalence of Bluetooth technology (BT) devices reflects the current IoT penetration in a number of digital products such as mobile phones, car hands-free sets, navigation systems, etc. BT devices emit signals with a unique Media Access Identification (MAC-ID) number that can be read by BT sensors within the coverage area. Readers placed at different locations can be used to identify the movement of the devices. Complemented by other data sources such as traffic signals, or bus GPS, research problems that can be addressed include vehicle travel time on motorway and arterial streets, dynamic (time dependent) O-D matrices on the network, identification of critical intersections, and accurate and reliable real time transport network state information [39]. There are many privacy concerns by such usages and digital forgetting is an emerging domain of research in IoT where privacy is a concern [44].

Another important application in mobile IoT domain is efficient logistics management [39]. This includes monitoring the items being transported as well as efficient transportation planning. The monitoring of items is carried out more locally, say, within a truck replicating enterprise domain but transport planning is carried out using a large scale IoT network.

## 5. Cloud centric Internet of Things

The vision of IoT can be seen from two perspectives – 'Internet' centric and 'Thing' centric. The Internet centric architecture will involve internet services being the main focus while data is contributed by the objects. In the object centric architecture [45], the smart objects take the center stage. In our work, we develop an Internet centric approach. A conceptual framework integrating the ubiquitous sensing devices and the applications is shown in Figure 4. In order to realize the full potential of cloud computing as well as ubiquitous Sensing, a combined framework with a cloud at the center seems to be most viable. This not only gives the flexibility of dividing associated costs in the most logical manner but is also highly scalable. Sensing service providers can join the network and offer their data using a storage cloud; analytic tool developers can provide their software tools; artificial intelligence experts can provide their data mining and machine learning tools useful in converting information to knowledge and finally computer graphics designer can offer a variety of visualization tools. The cloud computing can offer these services as Infrastructures, Platforms or Software where the full potential of human creativity can be tapped using them as services. This in some sense agrees with the ubicomp vision of Weiser as well as Rogers human centric approach. The data generated, tools used and the visualization created *disappears* into the background, tapping the full potential of the Internet of Things in various application domains. As can be seen from Figure 4, the Cloud integrates all ends of ubicomp by providing scalable storage, computation time and other tools to build new businesses. In this section we describe the cloud platform using Manjrasoft Aneka and Microsoft Azure platforms to demonstrate how cloud integrates storage, computation and visualization paradigms. Furthermore, we introduce an important realm of interaction between cloud which is useful for combining public and private clouds using Aneka. This interaction is critical for application developers in order to bring sensed information, analytics algorithms and visualization under one single seamless framework.

### 5.1. Aneka cloud computing platform

Aneka is a .NET-based application development Platform-as-a-Service (PaaS), which can utilize storage and compute resources of both public and private clouds [46]. It offers a runtime environment and a set of APIs that enable developers to build customized applications by using multiple programming models such as Task Programming, Thread Programming and MapReduce Programming. Aneka provides a number of services that allow users to control, auto-scale, reserve, monitor and bill users for the resources used by their applications. In the context of Smart Environment application, Aneka PaaS has another important characteristic of supporting the provisioning of resources on public clouds such as Microsoft Azure, Amazon EC2, and GoGrid, while also harnessing private cloud resources ranging from desktops and clusters, to virtual datacenters. An overview of Aneka PaaS is shown in Figure 5. For the application developer, the cloud service as well as ubiquitous sensor data is hidden and they are provided as services at a cost by the Aneka provisioning tool.

Automatic management of clouds for hosting and delivering IoT services as SaaS (Software-as-a-Service) applications will be the integrating platform of the Future Internet. There is a need to create data and service sharing infrastructure which can be used for addressing several application scenarios. For example, anomaly detection in sensed data carried out at the Application layer is a service which can be shared between several applications. Existing/new applications deployed as a hosted service and accessed over the Internet is referred to as SaaS. To manage SaaS applications on a large scale, the Platform as a Service (PaaS) layer needs to coordinate the cloud (resource provisioning and application scheduling) without impacting the Quality of Service (QoS) requirements of any application. The autonomic management components are to be put in place to schedule and provision resources with a higher level of accuracy to support IoT applications. This coordination requires the PaaS layer to support autonomic management capabilities required to handle the scheduling of applications and resource provisioning such that the user QoS requirements are satisfied. The autonomic management components are thus put in place to schedule and provision resources with a higher level of accuracy to support IoT applications. The autonomic management system will tightly integrate the following services with the Aneka framework: Accounting, Monitoring and Profiling, Scheduling, and Dynamic Provisioning. Accounting, Monitoring, and Profiling will feed the sensors of the autonomic manager, while the managers effectors will control Scheduling and Dynamic Provisioning. From a logical point of view the two components that will mostly take advantage from the introduction of autonomic features in Aneka are the application scheduler and the dynamic resource provisioning.

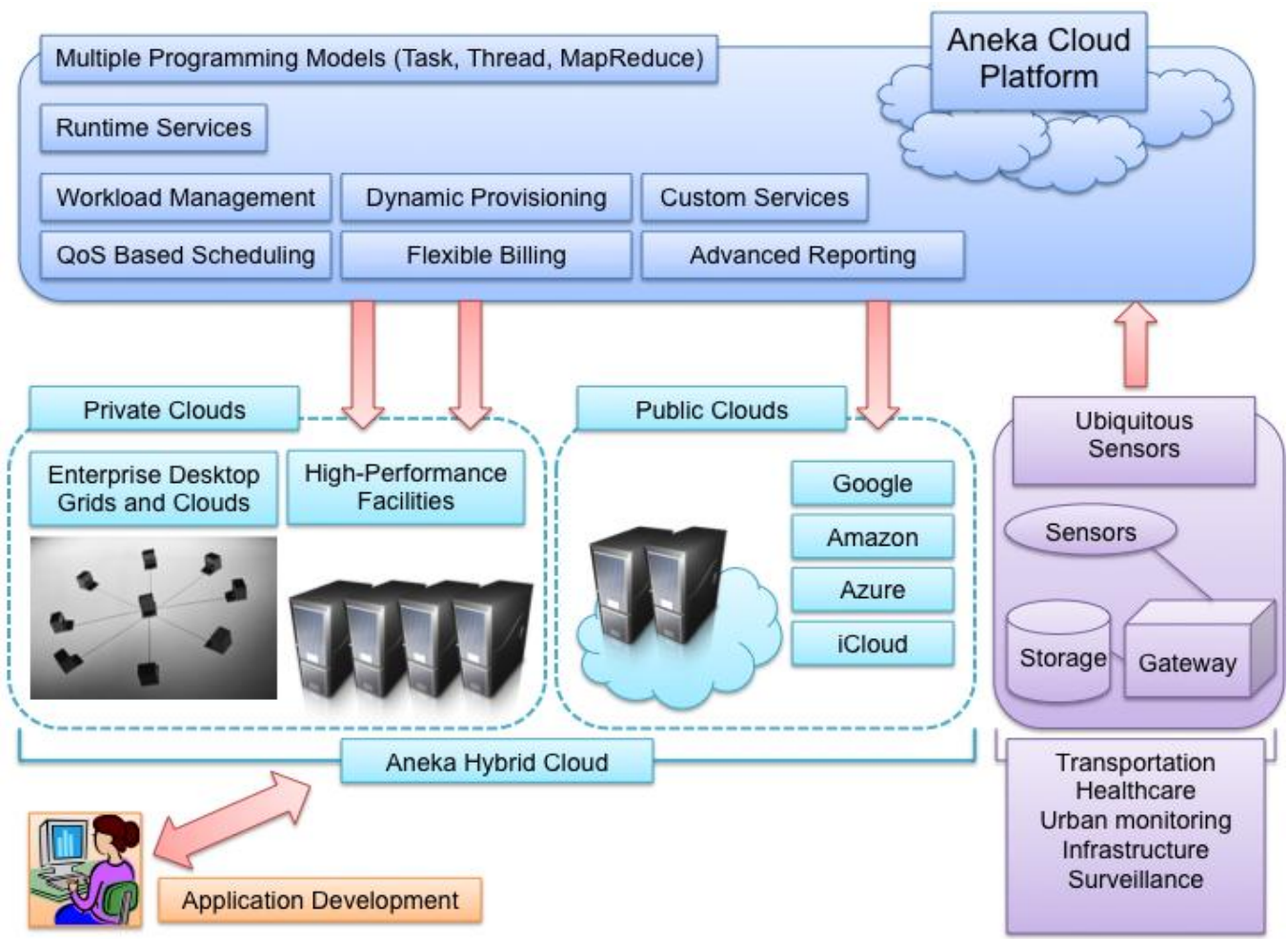

Figure 5: Overview of Aneka within Internet of Things Architecture [45]

*5.2. Application scheduler and Dynamic Resource Provisioning in Aneka for IoT applications*

The Aneka scheduler is responsible for assigning each resource to a task in an application for execution based on user QoS parameters and the overall cost for the service provider. Depending on the computation and data requirements of each Sensor Application, it directs the dynamic resource-provisioning component to instantiate or terminate a specified number of computing, storage, and network resources while maintaining a queue of tasks to be scheduled. This logic is embedded as multi-objective application scheduling algorithms. The scheduler is able to mange resource failures by reallocting those tasks to other suitable Cloud resources.

The Dynamic Resource Provisioning component implements the logic for provisioning and managing virtualised resources in the private and public cloud computing environments based on the resource requirements as directed by the application scheduler. This is achieved by dynamically negotiating with the Cloud Infrastructure as a Service (IaaS) providers for the right kind of resource for a certain time and cost by taking into account the past execution history of applications and budget availability. This decision is made at run-time, when SaaS applications continuously send requests to the Aneka cloud platform [47].

**6. IoT Sensor Data Analytics SaaS using Aneka and Microsoft Azure**

Microsoft Azure is a cloud platform, offered by Microsoft, includes four components as summarized in [46]. There are several advantages for integrating Azure and Aneka. Aneka can launch any number of instances on the Azure cloud to run their applications. Essentially, it provides the provisioning infrastructure. Similarly, Aneka provides advanced PaaS features as shown in Figure 5. It provides multiple programming models (Task, Thread, MapReduce), runtime execution services, workload management services, dynamic provisioning, QoS based scheduling and flexible billing.

Table 3: Microsoft Azure Components

| Microsoft Azure | On demand compute services, Storage services |
|---|---|
| SQL Azure | Supports Transact-SQL and support for the synchronization of relational data across SQL Azure and on-premises SQL Server |
| AppFabric | Interconnecting cloud and on-premise applications; Accessed through the HTTP REST API |
| Azure Marketplace | Online service for making transactions on Apps and Data |

As discussed earlier, trealizese thubicompmp vision, tools and data needs to be shared between application developers to create new apps. There are two major hurdles in such an implementation. Firstly, interaction between clouds becomes critical which is addressed by Aneka in the InterCloud model. Aneka support for InterCloud model enables the creation of a hybrid Cloud computing environment which combines the resources of private and public Clouds. That is, whenever private Cloud is unable to meet application QoS requirements, Aneka leases extra capability from public Cloud to ensure that application is able to execute within a specified deadline in a seamless manner [47]. Secondly, data analytics and artificial intelligence tools are computationally demanding which requires huge resources. For data analytics and artificial intelligence tools, the Aneka task programming model provides the ability of expressing applications as a collection of independent tasks. Each task can perform different operations, or the same operation on different data, and can be executed in any order by the runtime environment. In order to demonstrate this, we have used a scenario where there are multiple analytics algorithm and multiple data sources. A schematic of the interaction between Aneka and Azure is given in Figure 6 where Aneka Worker Containers are deployed as instances of Azure Worker Role. The Aneka Master Container will be deployed in the on-premises private cloud, while Aneka Worker Containers will be run as instances of Microsoft Azure Worker Role. As shown in the Figure 6, there are two types of Microsoft Azure Worker Roles used. These are the Aneka Worker Role and Message Proxy Role. In this case, one instance of the Message Proxy Role and at least one instance of the Aneka Worker Role are deployed. The maximum number of instances of the Aneka Worker Role that can be launched is limited by the subscription offer of Microsoft Azure Service that a user selects. In this deployment scenario, when a user submits an application to the Aneka Master, the job units will be scheduled by the Aneka Master by leveraging on-premises Aneka Workers, if they exist, and Aneka Worker instances on Microsoft Azure simultaneously. When Aneka Workers finish the execution of Aneka work units, they will send the results

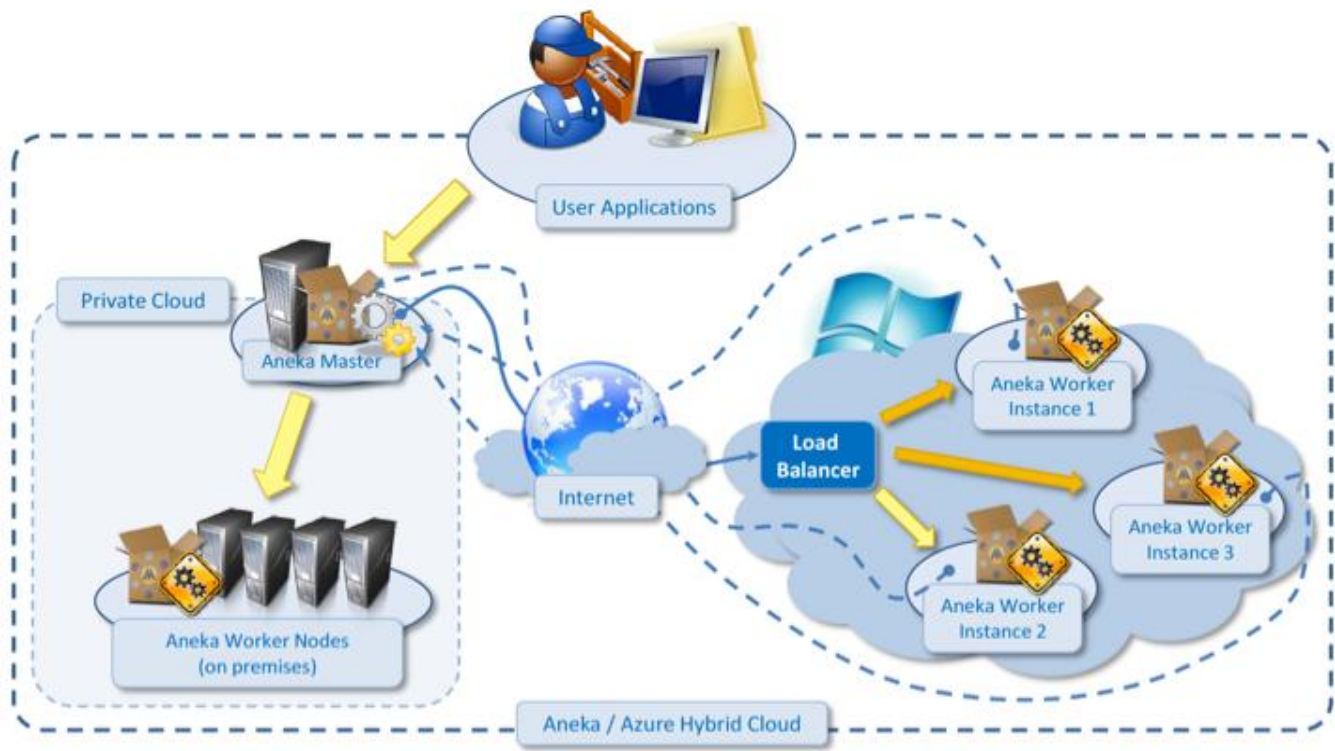

Figure 6: Schematic of Aneka/Azure Interaction for data analytics application [45].

back to Aneka Master, and then Aneka Master will send the result back to the user application.

Another important feature required for seamless independent IoT working architecture is SaaS to be updated by the developers dynamically. In this example, analytics tools (usually in the form of DLLs) have to be updated and used by several clients. Due to administrative privileges provided by Azure, this becomes a non-trivial task. Management Extensibility Framework (MEF) provides a simple solution to the problem. The MEF is a composition layer for .NET that improves the flexibility, maintainability and testability of large applications. MEF can be used for third-party plugin, or it can bring the benefits of a loosely-coupled plugin-like architecture for regular applications. It is a library for creating lightweight, extensible applications. It allows application developers to discover and use extensions with no configuration required. It also lets extension developers easily encapsulate code and avoid fragile hard dependencies. MEF not only allows extensions to be reused within applications, but across applications as well. MEF provides a standard way for the host application to expose itself and consume external extensions. Extensions, by their nature, can be reused amongst different applications. However, an extension could still be implemented in a way that is application-specific. The extensions themselves can depend on one another and MEF will make sure they are wired together in the correct order. One of the key design goals of IoT web application is, it would be extensible and MEF provides this solution. With MEF we can use different algorithms (as and when it becomes available) for IoT data analytics: e.g. drop an analytics assembly into a folder and it instantly becomes available to the application. The system context diagram of the developed data analytics is given in Figure 7.

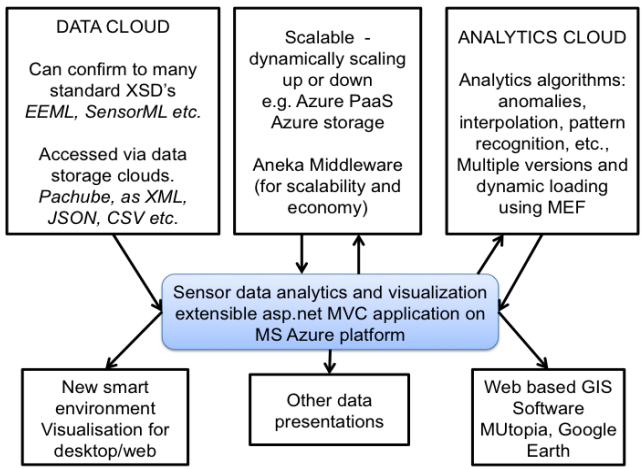

Figure 7: System Context Diagram

## 7. Open Challenges and Future Directions

The proposed Cloud centric vision comprises of a flexible and open architecture that is user centric and enables different players to interact in the IoT framework. It allows interaction in a manner suitable for their own requirements, rather than the IoT being thrust upon them. In this way, the framework includes provisions to meet different requirements for data ownership, security, privacy, and sharing of information.

Some open challenges are discussed based on the IoT elements presented earlier. The challenges include IoT specific challenges such as privacy, participatory sensing, data analytics, GIS based visualization and Cloud computing apart from the standard WSN challenges including architecture, energy efficiency, security, protocols, and Quality of Service. The end goal is to have Plug n' Play smart objects which can be deployed in any environment with an interoperable backbone allowing them to blend with other smart objects around them. Standardization of frequency bands and protocols plays a pivotal role in accomplishing this goal.

A roadmap of key developments in IoT research in the context of pervasive applications is shown in Figure 8, which includes the technology drivers and key application outcomes expected in the next decade [8]. The section ends with a few international initiatives in the domain which could play a vital role in the success of this rapidly emerging technology.

*7.1. Architecture*

Overall architecture followed at the initial stages of IoT research will have a severe bearing on the field itself and needs to be investigated. Most of the work relating to IoT architecture has been from the wireless sensor networks perspective [48]. European Union projects of SENSEI [49] and Internet of Things-Architecture (IoT-A) [50] have been addressing the challenges particularly from WSN perspective and have been very successful for defining the architecture for different applications. We are referring architecture to overall IoT where the user is at the center and will enable the use of data and infrastructure to develop new applications. An architecture based on cloud computing at the center has been proposed in this paper. However, this may not be the best option for every application domain particularly for defense where human intelligence is relied upon. Although we see cloud centric architecture to be the best where cost based services are required, other architectures should be investigated for different application domains.

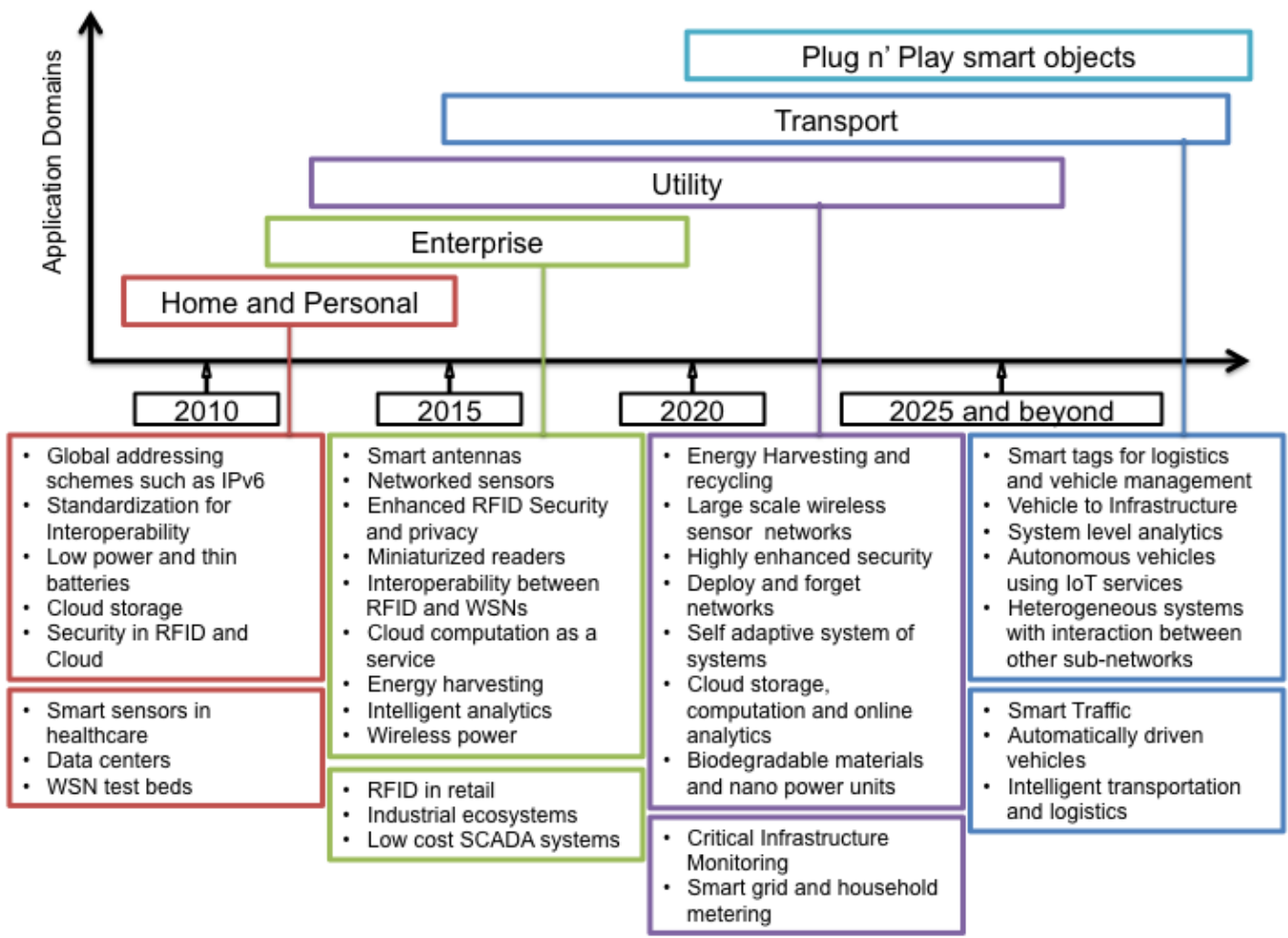

Figure 8: Roadmap of key technological developments in the context of IoT application domains envisioned

## 7.2. Energy efficient sensing

Efficient heterogeneous sensing of the urban environment needs to simultaneously meet competing demands of multiple sensing modalities. This has implications on network traffic, data storage and energy utilization. Importantly, this encompasses both fixed and mobile sensing infrastructure [51] as well as continuous and random sampling. A generalized framework is required for data collection and modelling that effectively exploits spatial and temporal characteristics of the data, both in the sensing domain as well as the associated transform domains. For example, urban noise mapping needs an uninterrupted collection of noise levels using battery powered nodes using fixed infrastructure and participatory sensing [51] as a key component for health and quality of life services for its inhabitants.

Compressive sensing enables reduced signal measurements without impacting accurate reconstruction of the signal. A signal sparse in one basis may be recovered from a small number of projections onto a second basis that is incoherent with the first [52]. The problem reduces to finding sparse solutions through smallest $l1$-norm coefficient vector that agrees with the measurements. In the ubiquitous sensing context, this has implications for data compression, network traffic and the distribution of sensors. Compressive wireless sensing (CWS) utilizes synchronous communication to reduce the transmission power of each sensor [53]; transmitting noisy projections of data samples to a central location for aggregation.

## 7.3. Secure reprogrammable networks and Privacy

Security will be a major concern wherever networks are deployed at large scale. There can be many ways the system could be attacked - disabling the network availability; pushing erroneous data into the network; accessing personal information; etc. The three physical components of IoT - RFID, WSN and cloud are vulnerable to such attacks. Security is critical to any network [54,55] and the first line of defence against data corruption is cryptography.

Of the three, RFID (particularly passive) seems to be the most vulnerable as it allows person tracking as well as the objects and no high level intelligence can be enabled on these devices [15]. These complex problems however have solutions which can be provided using cryptographic methods and deserves more research before they are widely accepted.

Against outsider attackers, encryption ensures data confidentiality, whereas message authentication codes ensure data integrity and authenticity [56]. Encryption, however, does not protect against insider malicious attacks, to address which non-cryptographic means are needed, particularly in WSNs. Also, periodically, new sensor applications need to be installed, or existing ones need to be updated. This is done by remote wireless reprogramming of all nodes in the network. Traditional network reprogramming consists solely of a data dissemination protocol that distributes code to all the nodes in the network without authentication, which is a security threat. A secure reprogramming protocol allows the nodes to authenticate every code update and prevent malicious installation. Most such protocols (e.g., [57]) are based on the benchmark protocol Deluge [58]. We need cryptographic add-ons to Deluge which lays foundation for more sophisticated algorithms to be developed.

Security in the cloud is another important area of research which will need more attention. Along with the presence of the data and tools, cloud also handles economics of IoT which will make it a bigger threat from attackers. Security and identity protection becomes critical in hybrid clouds where a private as well as public clouds will be used by businesses [59].

Remembering forever in the context of IoT raises many privacy issues as the data collected can be used in positive (for advertisement services) and negative ways (for defamation). Digital forgetting could emerge as one of the key areas of research to address the concerns and the development of appropriate framework to protect personal data [44].

*7.4. Quality of Service*

Heterogeneous networks are (by default) multi-service; providing more than one distinct application or service. This implies not only multiple traffic types within the network, but also the ability of a single network to support all applications without QoS compromise [60]. There are two application classes: throughput and delay tolerant elastic traffic of (e.g. monitoring weather parameters at low sampling rates), and the bandwidth and delay sensitive inelastic (real-time) traffic (e.g. noise or traffic monitoring), which can be further discriminated by data-related applications (e.g. high-vs.-low resolution videos) with different QoS requirements. Therefore, a controlled, optimal approach to serve different network traffics, each with its own application QoS needs is required [61]. It is not easy to provide QoS guarantees in wireless networks, as segments often constitute 'gaps' in resource guarantee due to resource allocation and management ability constraints in shared wireless media. Quality of Service in cloud Computing is another major research area which will require more and more attention as the data and tools become available on clouds. Dynamic scheduling and resource allocation algorithms based on particle swarm optimization are being developed. For high capacity applications and as IoT grows, this could become a bottleneck.

*7.5. New protocols*

The protocols at the sensing end of IoT will play a key role in complete realisation. They form the backbone for the data tunnel between sensors and the outer world. For the system to work efficiently, and energy efficient MAC protocol and appropriate routing protocol are critical. Several MAC protocols have been proposed for various domains with TDMA (collision free), CSMA (low traffic efficiency) and FDMA (collision free but requires additional circuitry in nodes) schemes available to the user [67]. None of them are accepted as a standard and with more 'things' available this scenario is going to get more cluttered which requires further research.

An individual sensor can drop out for a number of reasons, so the network must be self-adapting and allow for multi-path routing. Multi-hop routing protocols are used in mobile ad hoc networks and terrestrial WSNs [68]. They are mainly divided into three categories - data centric, location based and hierarchical, again based on different application domains. Energy is the main consideration for the existing routing protocols. In the case of IoT, it should be noted that a backbone will be available and the number of hops in the multi-hop scenario will be limited. In such a scenario, the existing routing protocols should suffice in practical implementation with minor modifications.

*7.6. Participatory Sensing*

A number of projects have begun to address the development of people centric (or participatory) sensing platforms [51,62-64]. As noted earlier, people centric sensing offers the possibility of low cost sensing of the environment localized to the user. It can therefore give the closest indication of environmental parameters experienced by the user. It has been noted that environmental data collected by user forms a social currency [65]. This results in more timely data being generated compared to the data available through a fixed infrastructure sensor network. Most importantly, it is the opportunity for the user to provide feedback on their experience of a given environmental parameter that offers valuable information in the form of context associated with a given event.

The limitations of people centric sensing place new significance on the reference data role provided by a fixed infrastructure IoT as a backbone. The problem of missing samples is a fundamental limitation of people centric sensing. Relying on users volunteering data and on the inconsistent gathering of samples obtained across varying times and varying locations (based on a user's desired participation and given location or travel path), limits the ability to produce meaningful data for any applications and policy decisions. Only in addressing issues and implications of data ownership, privacy and appropriate participation incentives, can such a platform achieve genuine end-user engagement. Further sensing modalities can be obtained through the addition of sensor modules attached to the phone for application specific sensing, such as air quality sensors [66] or biometric sensors. In such scenarios, smart phones become critical IoT nodes which are connected to the cloud on one end and several sensors at the other end.

*7.7. Data mining*

Extracting useful information from a complex sensing environment at different spatial and temporal resolutions is a challenging research problem in artificial intelligence. Current state-of-the-art methods use shallow learning methods where pre-defined events and data anomalies are extracted using supervised and unsupervised learning [69]. The next level of learning involves inferring local activities by using temporal information of events extracted from shallow learning. The ultimate vision will be to detect complex events based on larger spatial and longer temporal scales based on the two levels before. The fundamental research problem that arises in complex sensing environments of this nature is how to simultaneously learn representations of events and activities at multiple levels of complexity (i.e., events, local activities and complex activities). An emerging focus in machine learning research has been the field of deep learning [70], which aims to learn multiple layers of abstraction that can be used to interpret given data. Furthermore, the resource constraints in sensor networks create novel challenges for deep learning in terms of the need for adaptive, distributed and incremental learning techniques.

*7.8. GIS based visualization*

As new display technologies emerge, creative visualization will be enabled. The evolution from CRT to Plasma, LCD, LED, and AMOLED displays have given rise to highly efficient data representation (using touch interface) with the user being able to navigate the data better than ever before. With emerging 3D displays, this area is certain to have more research and development opportunities. However, the data which comes out of ubiquitous computing is not always ready for direct consumption using visualization platforms and requires further processing. The scenario becomes very complex for heterogeneous spatio-temporal data [71]. New visualization schemes for representation of heterogeneous sensors in 3D landscape that varies temporally have to be developed [72]. Another challenge of visualizing data collected within IoT is that they are geo-related and are sparsely distributed. To cope with such a challenge, a framework based on Internet GIS is required.

*7.9. Cloud Computing*

An integrated IoT and Cloud computing applications enabling the creation of smart environments such as Smart Cities need to be able to (a) combine services offered by multiple stakeholders and (b) scale to support a large number of users in a reliable and decentralized manner. They need to be able operate in both wired and wireless network environments and deal with constraints such as access devices or data sources with limited power and unreliable connectivity. The Cloud application platforms need to be enhanced to support (a) the rapid creation of applications by providing domain specific programming tools and environments and (b) seamless execution of applications harnessing capabilities of multiple dynamic and heterogeneous resources to meet quality of service requirements of diverse users.

The Cloud resource management and scheduling system should be able to dynamically prioritize requests and provision resources such that critical requests are served in real time. To deliver results in a reliable manner, the scheduler needs to be augmented with task duplication algorithms for failure management. Specifically, the Cloud application scheduling algorithms need to exhibit the following capability:

1. Multi-objective optimization: The scheduling algorithms should be able to deal with QoS parameters such as response time, cost of service usage, maximum number of resources available per unit price, and penalties for service degradation.
2. Task duplication based fault tolerance: Critical tasks of an application will be transparently replicated and executed on different resources so that if one resource fails to complete the task, the replicated version can be used. This logic is crucial in real-time tasks that need to be processed to deliver services in a timely manner.

*7.10. International Activities*

Internet of Things activities is gathering momentum around the world, with numerous initiatives underway across industry, academia and various levels of government, as key stakeholders seek to map a way forward for the coordinated realization of this technological evolution. In Europe, substantial effort is underway to consolidate the cross-domain activities of research groups and organizations, spanning M2M, WSN and RFID into a unified IoT framework. Supported by the European Commission 7th Framework program (EU-FP7) this includes the Internet of Things European Research Cluster (IERC). Encompassing a number of EU FP7 projects, its objectives are: to establish a cooperation platform and research vision for IoT activities in Europe and become a contact point for IoT research in the world. It includes projects such as CASAGRAS2, a consortium of international partners from Europe, the USA, China, Japan and Korea exploring issues surrounding RFID and its role in realizing the Internet of Things. As well, IERC includes the Internet of Things Architecture (IoT-A) project established to determine an architectural reference model for the interoperability of Internet-of-Things systems and key building blocks to achieve this. At the same time, the IoT Initiative (IoT-i) is a coordinated action established to support the development of the European IoT community. The IoT-i project brings together a consortium of partners to create a joint strategic and technical vision for the IoT in Europe that encompasses the currently fragmented sectors of the IoT domain holistically. Simultaneously, the SmartSantander project is developing a city scale IoT testbed for research and service provision deployed across the city of Santander, Spain, as well as sites located in the UK, Germany, Serbia and Australia.

At the same time large scale initiatives are underway in Japan, Korea, the USA and Australia, where industry, associated organizations and government departments are collaborating on various programs, advancing related capabilities towards an IoT. This includes smart city initiatives, smart grid programs incorporating smart metering technologies and roll-out of high speed broadband infrastructure. A continuing development of RFID related technologies by industry and consortiums such as the Auto-ID lab (founded at MIT and now with satellite labs at leading universities in South Korea, China, Japan, United Kingdom, Australia and Switzerland) dedicated to creating the Internet of Things using RFID and Wireless Sensor Networks are being pursued. Significantly, the need for consensus around IoT technical issues has seen the establishment of the Internet Protocol for Smart Objects (IPSO) Alliance, now with more than 60 member companies from leading technology, communications and energy companies, working with standards bodies, such as IETF, IEEE and ITU to specify new IP-based technologies and promote industry consensus for assembling the parts for the Internet of Things. Substantial IoT development activity is also underway in China, with its 12th Five Year Plan (2011-2015), specifying IoT investment and development to be focused on: smart grid; intelligent transportation; smart logistics; smart home; environment and safety testing; industrial control and automation; health care; fine agriculture; finance and service; military defense.

This is being aided by the establishment of an Internet of Things center in Shanghai (with a total investment over US$ 100million) to study technologies and industrial standards. An industry fund for Internet of Things, and an Internet of Things Union 'Sensing China' has been founded in Wuxi, initiated by more than 60 telecom operators, institutes and companies who are the primary drivers of the industry.

## 8. Summary and Conclusions

The proliferation of devices with communicating-actuating capabilities is bringing closer the vision of an Internet of Things, where the sensing and actuation functions seamlessly blend into the background and new capabilities are made possible through access of rich new information sources. The evolution of the next generation mobile system will depend on the creativity of the users in designing new applications. IoT is an ideal emerging technology to influence this domain by providing new evolving data and the required computational resources for creating revolutionary apps.

Presented here is a user-centric cloud based model for approaching this goal through the interaction of private and public clouds. In this manner, the needs of the end-user are brought to the fore. Allowing for the necessary flexibility to meet the diverse and sometimes competing needs of different sectors, we propose a framework enabled by a scalable cloud to provide the capacity to utilize the IoT. The framework allows networking, computation, storage and visualization themes separate thereby allowing independent growth in every sector but complementing each other in a shared environment. The standardization which is underway in each of these themes will not be adversely affected with Cloud at its center. In proposing the new framework associated challenges have been highlighted ranging from appropriate interpretation and visualization of the vast amounts of data, through to the privacy, security and data management issues that must underpin such a platform in order for it to be genuinely viable. The consolidation of international initiatives is quite clearly accelerating progress towards an IoT, providing an overarching view for the integration and functional elements that can deliver an operational IoT.

**Acknowledgements**

There have been many contributors for this to take shape and the authors are thankful to each of them. We specifically would like to thank Mr. Kumaraswamy Krishnakumar, Dr. Jiong Jin, Dr. Yee Wei Law, Prof. Mike Taylor, Mr. Aravinda Rao and Prof. Chris Leckie. The work is partially supported by Australian Research

Council's LIEF (LE120100129), Linkage grants (LP120100529) and Research Network ion Intelligent Sensors, Sensor networks and Information Processing (ISSNIP). The authors are participants in European 7thFramework projects on Smart Santander and Internet of Things - Initiative and are thankful for their support.